\documentclass{jps-cp}
\usepackage{txfonts} 
\newcommand{\Rm}{R_\text{m}}
\newcommand{\Tm}{T_\text{m}}
\newcommand{\so}{\mathcal{S}_0}
\usepackage{braket}
\usepackage{xcolor}

\title{Impact of Channel Mixing on the Visibility of Two-particle Interferometry in Quantum Hall Edge States
}

\author{Matteo \textsc{Acciai}$^{1}$, Preden \textsc{Roulleau}$^{2}$, D. Christian \textsc{Glattli}$^{2}$ and Janine \textsc{Splettstoesser}$^{1}$}

\inst{$^{1}$Department of Microtechnology and Nanoscience (MC2), Chalmers University of Technology, S-412 96 G\"oteborg, Sweden \\
$^{2}$Universit\'e Paris-Saclay, CEA, CNRS, SPEC, 91191 Gif-sur-Yvette, France
}

\email{acciai@chalmers.se}

\abst{
We consider a two-particle interferometer, where voltage sources applied to ohmic contacts inject electronic excitations into a pair of copropagating edge channels. We analyze the impact of channel mixing due to inter-edge tunneling on the current noise measured at the output of the interferometer. Due to
this mixing, the noise suppression typically expected for synchronized injecting sources is incomplete, thereby reducing the visibility of the interference. We investigate to which extent the impact of mixing on the noise visibility depends on different shapes of the voltage drives. 
Furthermore, we compare a simple model involving a single mixing point between the sources and the quantum point contact to the more realistic case of a continuous distribution of weak mixing points.}

\kword{fluctuations and noise, quantum Hall effect, two-particle interference}

\begin{document}
\maketitle

\section{Introduction}
In the last two decades, the research field known as electron quantum optics~\cite{Grenier2011,Bocquillon2013review,Bauerle2018} has emerged, following the technological advances that have led to the more and more accurate manipulation of electric currents, down to the single-electron level. In particular, the ability to implement reliable, on-demand single-electron sources~\cite{Feve2007,Hermelin2011,Dubois2013,Fletcher2013}, has allowed researchers to replicate famous quantum optic experiments, such as the Hong-Ou-Mandel (HOM) interference~\cite{Hong1987}, in electronic quantum
devices~\cite{Olkhovskaya2008,Jonckheere2012,Dubois2013,Bocquillon2013}.
The interest in electron quantum optics is due to multiple reasons. On the one hand, it offers the possibility to explore in a controlled framework fundamental differences due to particle statistics (fermions vs bosons and, more recently, even anyons~\cite{Carrega2021}), as well as the effect of many-body interactions. On the other hand, the coherent manipulation of single-electron states in condensed matter systems may pave the way to the realization of electronic flying qubits~\cite{Bauerle2018,Edlbauer2022Jul}, with potential applications to quantum information processing. Finally, generalizations of the HOM interference principle to ac voltage-driven contacts emitting coherent electron-hole pairs to probe the Hanbury Brown-Twiss phase, as proposed by Rychkov, Polianski and B\"uttiker \cite{Rychkov2005}, provide a convenient way to test the quantum coherence of carriers, as realized in a recent experiment \cite{Taktak2021}.

In view of these developments, it is crucial to have an accurate understanding of the ac regime, relevant for the operation of single-electron sources, and to identify detrimental mechanisms that may lead to the loss of coherence of the propagating electronic states.
In this paper, we address a two-particle interferometer in the integer quantum Hall regime.
In a previous work~\cite{Acciai2022PRB}, we have shown that channel mixing can be responsible for an incomplete suppression of the HOM noise, thereby reducing the interference visibility. Here, after reporting some of the main results from~\cite{Acciai2022PRB}, we investigate the features of the HOM interference curve and its associated visibility by comparing three different periodic drives. Furthermore, we extend our previous results by including
an analysis of the HOM noise in the presence of a continuous distribution of weak mixing points.
Even though, strictly speaking, HOM interference deals with single-particle states, throughout this paper we will use the same word to refer to the generalized interference occurring when the incoming states are composed of an arbitrary number of electron-hole pairs excited by ac voltage drives.

\begin{figure}[t]
\centering
\includegraphics[width=0.5\textwidth]{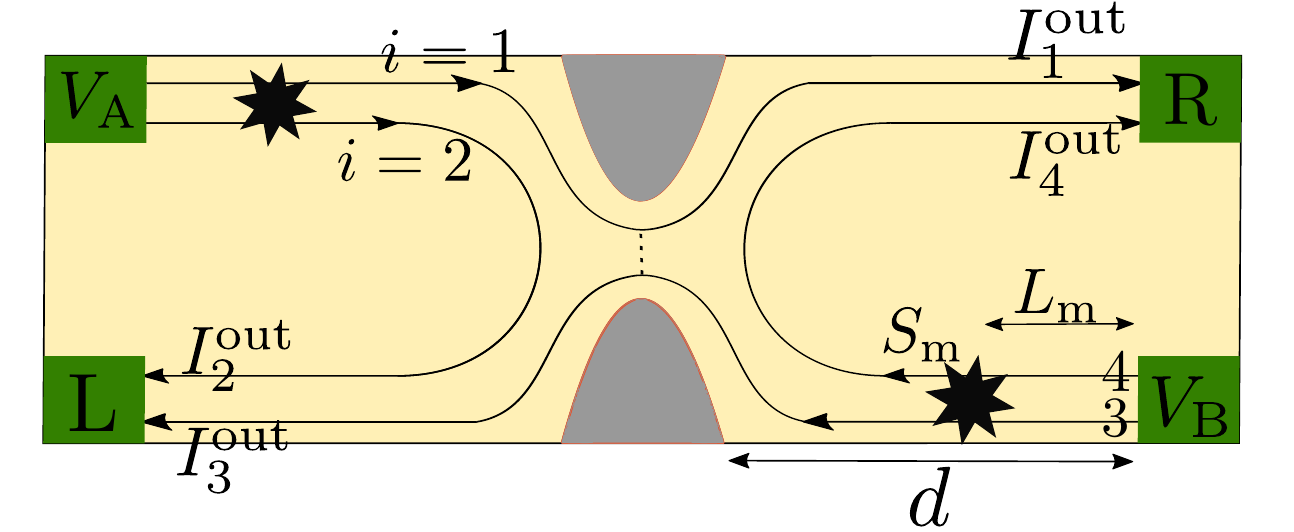}
\caption{Sketch of an electronic ``collider'' in the integer quantum Hall regime. Electronic states are injected into the edge channels via voltage sources ($V_\text{A}$ and $V_\text{B}$) and propagate towards a central quantum point contact, acting as a beamsplitter.
Two black stars indicate the presence of possible mixing points, inducing inter-edge tunneling. We assume them to be symmetrically placed at distance $L_\text{m}$ from the sources and described by identical scattering matrices $S_\mathrm{m}$. Current fluctuations are measured at the output contacts (R and L).}
\label{f1}
\end{figure}

\section{HOM noise: model and basic results}
In this section, we briefly 
summarize the theoretical modeling of a HOM-like interferometer in the integer quantum Hall effect at filling factor $\nu=2$, as sketched in Fig.~\ref{f1} and inspired by experiments~\cite{Bocuillon2013Science,Taktak2021}: 
two pairs of edge channels are connected to ohmic contacts where ac drives $V_\text{A}(t)=V(t)$ and $V_\text{B}(t)=V(t+\delta)$ are applied, with a tunable time delay $\delta$.
It is convenient to write $V(t)=V_\text{dc}+V_\text{ac}(t)$, with $\int_{0}^{\mathcal{T}}V_\text{ac}(t)\,\mathrm{d}t=0$ and $\mathcal{T}=2\pi\Omega^{-1}$ the period of the drive. The electronic states generated by the driven contacts are sent to a central quantum point contact (QPC), acting as a beamsplitter. Since electron-electron interactions were found to be irrelevant for the HOM noise suppression in voltage-driven interferometers~\cite{Rebora2020,Acciai2022PRB,Safi2014,Safi2022}, we neglect them here and describe the system by a Floquet scattering theory 
approach~\cite{moskalets-book}.
The incoming state on channel $i=1,...,4$ can be decomposed as a superposition of plane waves
\begin{equation}
    \hat{\psi}_{i,\text{in}}(x,t)=\int_{-\infty}^{\infty}\frac{\mathrm{d} E}{\sqrt{h v_{i}}} \sum_{l\in\mathbb{Z}}p_{l}  \hat{a}_{i,\text{in}}(E-l\hbar\Omega)e^{-\frac{i E}{\hbar}\left(t - \frac{x}{v_{i}}\right)},
\end{equation}
with the annihilation operator $\hat{a}_{i,\text{in}}(E)$ at energy $E$ and channel velocity $v_i$. The 
Floquet amplitudes
\begin{equation}
    p_l=\frac{1}{\mathcal{T}}\int_{-\mathcal{T}/2}^{\mathcal{T}/2}\mathrm{d}t\,e^{il\Omega t}e^{\frac{ie}{\hbar}\int_{-\infty}^t \mathrm{d}t' V_\text{ac}(t')}
    \label{eq:pl}
\end{equation}
encode the effect of the ac drive, 
expressing the probability amplitude that $|l|$ quanta of energy $\hbar\Omega$ are absorbed $(l>0)$ or emitted $(l<0)$ by an electron. Finally, the QPC is described by an energy-independent scattering matrix, so that the outgoing operators read
\begin{equation}
    \begin{pmatrix}
    \hat{a}_{1,\text{out}}(E)\\
    \hat{a}_{3,\text{out}}(E)
    \end{pmatrix}
    =
    \begin{pmatrix}
    \sqrt{T} & i\sqrt{R}\\
    i\sqrt{R} & \sqrt{T}
    \end{pmatrix}
    \begin{pmatrix}
    \hat{a}_{1,\text{in}}(E)\\
    \hat{a}_{3,\text{in}}(E)
    \end{pmatrix},
    \qquad
    \begin{pmatrix}
    \hat{a}_{2,\text{out}}(E)\\
    \hat{a}_{4,\text{out}}(E)
    \end{pmatrix}
    =
    \begin{pmatrix}
    \hat{a}_{2,\text{in}}(E)\\
    \hat{a}_{4,\text{in}}(E)
    \end{pmatrix},
\end{equation}
where $T=1-R$ is the transmission probability at the QPC.
The zero-frequency noise associated with the outgoing charge currents $\hat{I}_{i,\text{out}}=-e\hat{\psi}^\dagger_{i,\text{out}}\hat{\psi}_{i,\text{out}}$ is
\begin{equation}
    \mathcal{S}_{ij}=2\int_{-\mathcal{T}/2}^{\mathcal{T}/2}\frac{\mathrm{d}t}{\mathcal{T}}\int_{-\infty}^{+\infty}\mathrm{d}t'\left[\Braket{\hat{I}_{i}^\text{out}\left(t+\frac{t'}{2}\right)\hat{I}_{j}^\text{out}\left(t-\frac{t'}{2}\right)}-\Braket{\hat{I}_{i}^\text{out}\left(t+\frac{t'}{2}\right)}\Braket{\hat{I}_{j}^\text{out}\left(t-\frac{t'}{2}\right)}\right]\,.
    \label{eq:noise_general_definition}
\end{equation}
The HOM noise is the cross-correlation noise of the outgoing currents collected at the output contacts, namely $\hat{I}_\text{R}^{\text{out}}=\hat{I}_{1}^{\text{out}}+\hat{I}_{4}^{\text{out}}$ and $\hat{I}_\text{L}^{\text{out}}=\hat{I}_{2}^{\text{out}}+\hat{I}_{3}^{\text{out}}$. It is also standard to subtract the equilibrium noise one would find when the sources are switched off, leading to the definition
\begin{equation}
    \mathcal{S}_\text{HOM}=-\left(\mathcal{S}_\text{RL}^\text{on,on}-\mathcal{S}_\text{RL}^\text{off,off}\right).
\end{equation}
A calculation of the HOM noise at finite temperature $\theta$ leads to the following result~\cite{Dubois2013PRB}
\begin{equation}
\mathcal{S}_\text{HOM}=RT\frac{e^2\Omega}{\pi}\sum_{l\in\mathbb{Z}} \Pi_l(\delta)\left[l\coth\left(\frac{l\hbar\Omega}{2k_\text{B}\theta}\right)-\frac{2k_\text{B}\theta}{\hbar\Omega}\right]\equiv RT\mathcal{S}_0(\delta),\qquad 
\Pi_l(\delta)=\Big|\sum_{m\in\mathbb{Z}} p_m p_{m-l}^* e^{im\Omega\delta}\Big|^2.
\label{eq:base-HOM}
\end{equation}
The HOM noise vanishes when the two ac sources are synchronized, $\mathcal{S}_0(\delta=0)=0$, as expected for indistinguishable incoming states. 
Interestingly, the same is true even when interactions are included~\cite{Rebora2020,Safi2014,Safi2022,Acciai2022PRB}, in contrast to what has been reported when the interfering states are injected with different schemes than the one we address here~\cite{Bocuillon2013Science,Wahl2014,Freulon2015,Marguerite2016}. 
As we explain below, channel mixing is a plausible cause of the incomplete suppression of the HOM noise in our setup.

\section{Reduced visibility by channel mixing}
We now introduce channel mixing due to inter-edge tunneling as a 
mechanism that can explain the recently observed reduction of visibility in the HOM noise~\cite{Taktak2021}. As sketched in Fig.~\ref{f1}, impurities arising from the native disorder in a two-dimensional electron gas may induce inter-edge tunneling along the co-propagating edge channels~\cite{Khaetskii_1992,Polyakov_1996,Pala_2005}.

\subsection{Symmetric interferometer}
We consider two mixing points, placed at a distance $L_\text{m}$ from the injecting sources. A more general configuration can be found in~\cite{Acciai2022PRB}. Moreover, we assume that
mixing occurs with equal probability $R_\text{m}$ in both input arms. Thus, the effect of mixing is encoded in the scattering matrix
\begin{equation}
    S_\text{m}=
    \begin{pmatrix}
    \sqrt{T_\text{m}} & i\sqrt{R_\text{m}}\\
    i\sqrt{R_\text{m}} & \sqrt{T_\text{m}}
    \end{pmatrix}\ ,
\end{equation}
connecting channels 1 and 2 (respectively 3 and 4).
Finally, we assume that the propagation velocities fulfill $v_1=v_3=v_\text{out}$ for the outer channels and $v_2=v_4=v_\text{in}$ for the inner ones. With these hypotheses, the HOM noise reads~\cite{Acciai2022PRB}
\begin{equation}
    \mathcal{S}_\text{HOM}(\Delta\tau,\delta)=2T^2\Rm\Tm\so(\Delta\tau)+RT(\Tm^2+\Rm^2)\so(\delta)+RT\Tm\Rm[\so(\delta+\Delta\tau)+\so(\delta-\Delta\tau)]\,,
    \label{eq:HOM_noise}
\end{equation}
where $\Delta\tau=|L_\text{m}(v_\text{in}^{-1}-v_\text{out}^{-1})|$.
This result can be 
straightforwardly interpreted by realizing that
the incoming excitations 
take different paths to reach the QPC due to mixing, thereby acquiring an effective time delay
on top of the bare time shift $\delta$ between the injecting sources. For instance, an excitation generated by $V_\text{A}$ on channel $2$ can be transferred to channel $1$ at the mixing point and eventually interfere at the QPC with another excitation injected by $V_\text{B}$ and propagating entirely on channel $3$. Such a situation results in an additional shift $\Delta\tau$,
as soon as the propagation velocities on inner and outer channels are different. 
Therefore, for nonvanishing mixing $\Rm\neq 0$ and $v_\text{in}\neq v_\text{out}$, the different terms in Eq.~\eqref{eq:HOM_noise} cannot simultaneously vanish, leading to an incomplete suppression of the HOM noise and a reduced visibility of the interference.

\subsection{Comparison between different drives}
\begin{figure}[t]
\centering
\includegraphics[width=\textwidth]{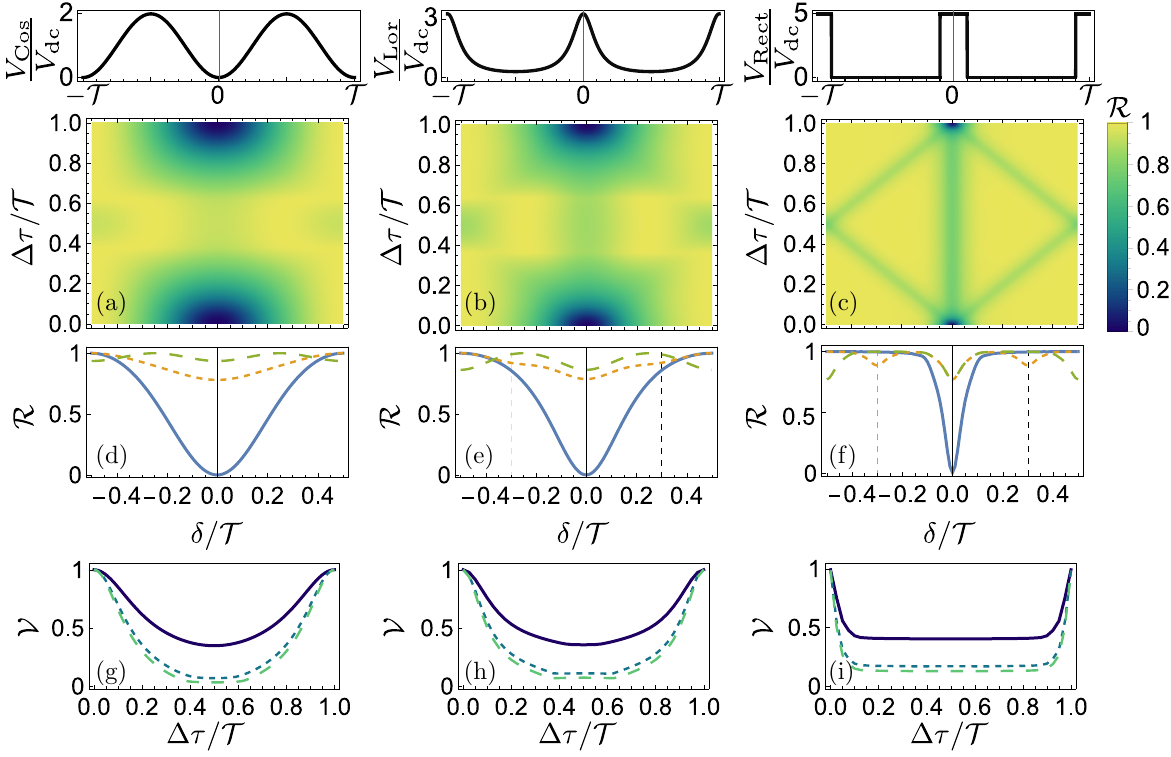}
\caption{Density plot of the normalized HOM noise $\mathcal{R}$ as a function of the time delay $\delta$ and the mixing-induced time shift $\Delta\tau$ for (a) cosine, (b) Lorentzian and (c) rectangular drives. In (b) and (c), the width-to-period ratio is fixed to $\eta=0.1$. Panels (d), (e), (f) show cuts of, respectively (a), (b), (c) at values $\Delta\tau/\mathcal{T}=0$ (thick line), $0.3$ (short dashed) and $0.5$ (long dashed). For all ac drives, the amplitude is fixed as $eV_\text{dc}=\hbar\Omega$. Other parameters in (a-f): $T=0.7$, $\Tm=0.5$, $k_\text{B}\theta=0.03\hbar\Omega$ (corresponding to roughly $25\,$mK at $\Omega/2\pi=14\,$GHz, which are the relevant experimental parameters in~\cite{Taktak2021}). Finally, panels (g), (h) and (i) show the visibility as a function of $\Delta\tau$ for, respectively, a cosine, a Lorentzian, and a rectangular drive. In these plots, we have also varied the mixing strength as $\Rm=0.1$ (full line), $0.3$ (short dashed) and $0.5$ (long dashed).}
\label{f2}
\end{figure}

In the following, we illustrate how the impact of channel mixing on the HOM visibility depends on the shape of the ac signal, by considering three different drives as shown in the upper row of Fig.~\ref{f2},
\begin{subequations}
\begin{align}
    V_\text{Cos}(t)&=V_\text{dc}[1-\cos(\Omega t)],\label{eq:drives_cos}\\
    V_\text{Lor}(t)&=\frac{V_\text{dc}}{\pi}\sum_{k\in\mathbb{Z}}\frac{\eta}{\eta^2+(t/\mathcal{T}-k)}=\frac{V_\text{dc}}{2}\frac{\sinh(2\pi\eta)}{\sinh^2(\pi\eta)+\sin^2(\pi t/\mathcal{T})},\\
    V_\text{Rect}(t)&=\frac{V_\text{dc}}{\eta}\sum_{k\in\mathbb{Z}}\Theta\left(\frac{\eta\mathcal{T}}{2}-\left|t-k\mathcal{T}\right|\right),
\end{align}
\label{eq:drives}
\end{subequations}
where $\Theta(x)$ is the Heaviside step function.
The first drive is a standard cosine signal. The second one is a periodic train of Lorentzian pulses, resulting in the injection of Levitons when $eV_\mathrm{dc}=\hbar\Omega$~\cite{Levitov1996,Dubois2013}; for this drive, $\eta$ represents the ratio between the full width at half maximum of each pulse and the period. The third signal is a rectangular drive with duty cycle $\eta$, reducing to a square wave for $\eta=1/2$. Clearly, $V_\text{Rect}$ and $V_\text{Lor}$ provide an additional knob compared to the cosine drive, as they allow one to vary the width of the pulse within each period.

In order to compare the HOM noises associated with different signals, it is convenient to study the parameter $\mathcal{R}$, namely $\mathcal{S}_\text{HOM}$ normalized with its maximum value as function of the detuning $\delta$, $\mathcal{S}_\text{max}=\text{max}_{\delta}\left\{\mathcal{S}_\text{HOM}(\delta)\right\}$, as well as its  visibility $\mathcal{V}$, with $\mathcal{S}_\text{min}=\text{min}_{\delta}\left\{\mathcal{S}_\text{HOM}(\delta)\right\}$, defined as,
\begin{equation}
\mathcal{R}=\frac{\mathcal{S}_\text{HOM}}{\mathcal{S}_\text{max}}
    \hspace{0.5cm} \text{and}     \hspace{0.5cm}
    \mathcal{V}= \frac{\mathcal{S}_\text{max}-\mathcal{S}_\text{min}}{\mathcal{S}_\text{max}+\mathcal{S}_\text{min}}.
    \label{eq:ratio}
\end{equation}
These dimensionless ratios are plotted in Fig.~\ref{f2} for the three drives in Eq.~\eqref{eq:drives}. In the density plots (a-c) we vary the intrinsic time delay $\delta$ between the voltage sources $V_\text{A}$ and $V_\text{B}$, as well as the mixing-induced time shift $\Delta\tau$. Note that a variation of the latter can be due to differences in the mixing point position $L_\text{m}$, but also in the velocity mismatch $v_\text{out}-v_\text{in}$. All the density plots in Fig.~\ref{f2} show that the HOM noise is fully suppressed at $\delta=0$ when $\Delta\tau=0$; then the dip is lifted by increasing $\Delta\tau$ and is eventually fully restored at $\Delta\tau=\mathcal{T}$. This restored dip,
which is clear from the analytical expression in Eq.~\eqref{eq:HOM_noise}, is physically due to the indistinguishability of electronic states emitted with a time delay of 
integer multiples of the drive period.

Besides this common feature, some differences emerge when we compare the Lorentzian and the rectangular drive with the cosine: as shown in Fig.~\ref{f2}(b,c,e,f), additional dips become visible
in the HOM noise, located around
$\delta=\pm\Delta\tau$. As we already noticed in~\cite{Acciai2022PRB}, this feature results from the collision between pulses that travelled along different edges at different velocities and appears when the pulse is narrow enough to distinguish the features from the main dip around $\delta=0$.
It is therefore clear that this effect is much more pronounced for the rectangular drive, as in this case the voltage drive abruptly shifts from $V_\text{dc}/\eta$ to $0$, the pulse being completely localized in a region of width $\eta$ within each period.

In panels (g,h,i) of Fig.~\ref{f2}, we plot the visibility as a function of $\Delta\tau$ for different mixing strengths $R_\text{m}$.
As is intuitively clear, a stronger
deterioration of the visibility is observed with increasing $R_\text{m}$.
The other important feature
is that the visibility is lost much more quickly with a rectangular drive, upon varying $\Delta\tau$. Once again, this behaviour can be understood as a consequence of the sharpness of the drive, which makes the minima of the different terms in Eq.~\eqref{eq:HOM_noise} well separated even for small $\Delta\tau$.

\subsection{Average over a continuous distribution of weak mixing points}
In a realistic sample, it is likely that multiple mixing points are present, calling for a more refined model to take this feature into account. In our previous work~\cite{Acciai2022PRB} we have addressed the case of two consecutive mixing points of arbitrary strength, showing that additional interference effects arise and that the visibility can both increase or decrease (compared with a single mixing point scenario), depending on the precise position of the scatterers along the edge channels. 
Here, we investigate the case of \emph{weak} mixing, meaning that the inter-edge tunneling probability $R_\text{m}$ is small. Under this condition, and considering that many mixing points may be present between the source and the QPC, it is reasonable to assume that a given excitation undergoes a single mixing event at a random position (depending on which 
impurity is responsible for the tunneling event).
Since noise measurements are the result of a time average, the 
detected signal will thus be equivalent to an average over many realizations of 
Eq.~\eqref{eq:HOM_noise}, where 
$\Delta\tau$ is randomly extracted according to the impurity distribution. Without any detailed knowledge of the sample, it is most reasonable to assume that mixing points are uniformly distributed in the interval $[0,L_\text{max}]$, with $L_\text{max}$ necessarily smaller than the distance $d$ between the source and the QPC. Thus, the averaged noise reads
\begin{equation}
\overline{\mathcal{S}_\text{HOM}(\delta)}=\frac{1}{\Delta\tau_\text{max}}\int_0^{\Delta\tau_\text{max}}\mathcal{S}_\text{HOM}(\Delta\tau,\delta)\,\text{d}(\Delta\tau).
\label{eq:hom_average}
\end{equation}

\begin{figure}[t]
\centering
\includegraphics[width=\textwidth]{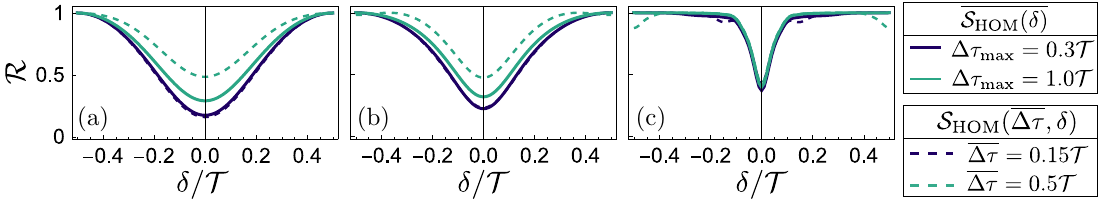}
\caption{Comparison between averaged noise of Eq.~\eqref{eq:hom_average} (full lines), normalized according to Eq.~\eqref{eq:ratio}, and the simple model with a single mixing point per arm (dashed lines). The full blue and green lines correspond to $\Delta\tau_\text{max}/\mathcal{T}=0.4$ and $1$, respectively. The dashed lines are obtained from Eq.~\eqref{eq:HOM_noise}, selecting the average time shift $\overline{\Delta\tau}=\Delta\tau_\text{max}/2$. Panels (a), (b) and (c) refer to a cosine, a Lorentzian, and a rectangular drive, respectively. All the other parameters are the same as in Fig.~\ref{f2}, while a weak mixing strength $R_\mathrm{m}=0.1$ is assumed here.}
\label{f3}
\end{figure}

We now illustrate how the presence of multiple mixing points 
modifies the results obtained from the simple model with a single mixing point on each input arm of the interferometer. 
For clarity,
we consider the low-temperature and low-amplitude regime, thus taking $eV_\text{ac}/\hbar\Omega\ll 1$ and $\theta\to 0$. Furthermore, we choose
the cosine drive, for which the photo-assisted probabilities are given in terms of the Bessel functions as $\Pi_l(\delta)=J_l^2(2eV_\text{ac}\sin(\Omega\delta/2)/\hbar\Omega)$. 
With this, we find
\begin{eqnarray}\mathcal{S}_\text{HOM}(\Delta\tau,\delta)\approx \frac{e^2\Omega}{\pi}2\left(\frac{eV_\text{ac}}{\hbar\Omega}\right)^2& \times&\Big\{2T^2{\Rm}{\Tm}\sin^2(\Omega\Delta\tau/2)
\label{eq:hom_low_sine}
\\ &&+RT(\Tm^2+\Rm^2)\sin^2(\Omega\delta/2)
+RT{\Tm}{\Rm}[1-\cos(\Omega\Delta\tau)\cos(\Omega\delta)]\Big\}.\nonumber
\end{eqnarray}
The advantage of the small $V_\text{ac}$ approximation is that the averaging can be performed analytically:
\begin{eqnarray}
\overline{\mathcal{S}_\text{HOM}(\delta)}\approx\frac{e^2\Omega}{\pi}2\left(\frac{eV_\text{ac}}{\hbar\Omega}\right)^2&\times&\Bigg\{2T^2{\Rm}{\Tm}\left[\frac{1}{2}-\frac{\sin(\Omega\Delta\tau_\text{max})}{2\Omega\Delta\tau_\text{max}}\right]\label{eq:hom_sine_av}
\\
&+&RT(\Tm^2+\Rm^2)\sin^2(\Omega\delta/2)
+RT{\Tm}{\Rm}\left[1-\frac{\cos(\Omega\delta)\sin(\Omega\Delta\tau_\text{max})}{\Omega\Delta\tau_\text{max}}\right]\Bigg\}.
\nonumber
\end{eqnarray}
Therefore, we can make a direct comparison between the result in the presence of multiple weak mixing points [Eq.~\eqref{eq:hom_sine_av}] and the 
result with just one mixing point in each input arm
[Eq.~\eqref{eq:hom_low_sine}]. 
First, we find by explicit substitution
\begin{equation}
\overline{\mathcal{S}_\text{HOM}(\delta)}\Big|_{\Omega\Delta\tau_\text{max}=n\pi}=\mathcal{S}_\text{HOM}\left(\frac{2m+1}{2}\frac{\pi}{\Omega},\delta\right),\forall n,m\in\mathbb{Z}.
\end{equation}
We find an even more direct relation 
when the mixing-induced time shifts are small compared to the period of the drive, namely $\Omega\Delta\tau\ll 1$ and $\Omega\Delta\tau_\text{max}\ll 1$. 
Expanding the sine/cosine functions,
we see that the 
terms in the second rows of Eqs.~\eqref{eq:hom_low_sine} and \eqref{eq:hom_sine_av} become identical, while the the ones in the first map onto each other if one replaces $\Delta\tau\leftrightarrow\Delta\tau_\text{max}/\sqrt{3}$. This shows that, in this regime, the results obtained in the presence or absence of multiple mixing points are essentially the same and a simple 
rescaling of $\Delta\tau$ maps them onto each other. For a generic drive and ac amplitude, no simple expression for $\overline{\mathcal{S}_\text{HOM}}$ can be found and one has to resort to a numerical evaluation of Eq.~\eqref{eq:hom_average}, see Fig.~\ref{f3}. The quantitative difference between $\mathcal{S}_\text{HOM}$ and $\overline{\mathcal{S}_\text{HOM}}$ crucially depends on $\Delta\tau_\text{max}$, which is linked to the sample size, as well as on the chosen drive. Nevertheless, the qualitative features of the refined model including multiple weak scatterers are the same as those observed in the simple two-point calculation, within the validity regime of weak tunneling for the continuous mixing-point model.

\section{Conclusion}
We have extended our previous analysis of the impact of tunneling-induced channel mixing on
the current noise in a HOM-like interferometer in the integer quantum Hall effect. 
Specifically, we have here
compared how the impact of mixing differs for different voltage drives. Furthermore, we have included the presence of multiple mixing points along the edge channels. Our results can be further tested by measuring the noise visibility as a function of the edge channel length. Such a measurement is expected to be within the reach of current experimental techniques.

\section*{Acknowledgements}
This work has received funding from the European Union's H2020 research and innovation program under grant agreement No.~862683.
Funding from the Knut and Alice Wallenberg Foundation through the Academy Fellow program (J.S. and M.A.) is also gratefully acknowledged. D.C.G. and P.R. thank funding by the French ANR contract FullyQuantum (ANR-16-CE30-0015).

\end{document}